\documentclass[10pt]{article}
\usepackage{graphicx}
\usepackage{url}
\usepackage{color}
\definecolor{light-gray}{gray}{0.95}
\definecolor{mygreen}{rgb}{0,0.6,0}
\definecolor{myamber}{rgb}{1.0, 0.49, 0.0}
\PassOptionsToPackage{hyperref}{color}
\usepackage{listings}
\usepackage{hyperref}

\begin{document}

% Page heads
\markboth{E. Givelberg}
{
Process-Oriented Parallel Programming
}

% Title portion

\title{
% A Large-scale Computation of a 3D Fourier Transform
%
Process-Oriented Parallel Programming
with an Application to Data-Intensive Computing
% Using Processes in a Data-intensive computation
% Object-Oriented Parallel Programming
}
\author{
Edward Givelberg
% \affil{Johns Hopkins University}
}

% \date{15 April 2014}

% \category{D.1.3}{Software}{Concurrent Programming}
% \category{D.1.4}{Software}{Object-oriented Programming}
% 
% \terms{
% Design, Languages
% }
% 
% \keywords{
% Parallel programming,
% programming languages,
% object-oriented programming,
% data-intensive computing
% }
% 
% \acmformat{
% E. Givelberg, 2014.
% Process-Oriented Parallel Programming
% with an Application to Data-Intensive Computing
% }
% 
% \begin{bottomstuff}
% This research was partially funded by a grant from Intel Corp.
% 
% Author's address: E. Givelberg,
% Department of Physics and Astronomy,
% Johns Hopkins University
% % givelberg@jhu.edu
% % Computer Science Department,
% % College of William and Mary; Y. Wu  {and} J. A. Stankovic,
% % Computer Science Department, University of Virginia; T. Yan,
% % Eaton Innovation Center; T. He, Computer Science Department,
% % University of Minnesota; C. Huang, Google; T. F. Abdelzaher,
% % (Current address) NASA Ames Research Center, Moffett Field, California 94035.
% \end{bottomstuff}
 
\maketitle

\begin{abstract}

We introduce
{\em process-oriented programming}
as a natural extension of object-oriented programming for parallel computing.
It is based on the observation that every class of an object-oriented
language can be instantiated as a process, accessible via a remote pointer.
The introduction of process pointers requires no syntax extension,
identifies processes with programming objects,
and enables 
processes to exchange information simply by executing remote methods.
% Syntactically, executing a method on a remote process is not different
% from method execution on an object.
% using process pointers.
% No new syntax is needed because this is just like object methods.
% just like object pointers, so no new syntax is needed.
% Process pointers enable processes to exchange information
% by executing remote methods
% Syntactically, executing a method on a remote process is not different
% from method execution on an object.
% The syntax extension is minimal.
% In C++, for example, it
% amounts to adding a parameter
% to the {\tt \color{blue} new} operator and
% introducing a single new keyword
% {\tt \color{blue} start}.
%
% The introduction of remote pointers
% % is a powerful syntactic device
% enables a straightforward extension of object-oriented
% programming languages to process-oriented programming
% with hardly any syntax additions.
%
% We argue that {\em object-based parallelism}
% is a high level abstraction, which is
% naturally suitable for reasoning about parallelism.
% Although shared memory and message passing can be realized
% in a process-oriented language,
% these are lower implementation-level models.
%
Process-oriented programming is a high-level language
% replacement
alternative
to multithreading, MPI and many other languages, environments
and tools currently used for parallel computations.
It
% offers the
implements
natural {\em object-based parallelism}
using only minimal syntax extension of existing languages,
such as C++ and Python,
% Potentially its most important impact is
% potentially leading to
and has therefore the potential to lead to
widespread adoption of parallel programming.
% as application developers realize the ability
% to easily create processes instead of using thread libraries,
% and
% to place different
% objects on different CPU cores.
We implemented a prototype system for running processes using
C++ with MPI
and used it to compute 
% We describe an application of process-oriented programming to
% the computation of
a large three-dimensional Fourier transform
on a computer cluster built of commodity hardware components.
% The process-oriented
% implementation of the Fourier transform
% was written using only a few hundred lines of code
% and
% translated into a C++/MPI application consisting of many thousands
% of lines of code.
% The computation of a
Three-dimensional Fourier transform 
% can be considered as
is
a prototype of a data-intensive application with a complex data-access
pattern.
The process-oriented code is only a few hundred lines long,
% but it translates into many thousands of lines of C++ with MPI.
% We demonstrate that the process-oriented application
and
% the resulting code
attains very high
data throughput by achieving massive parallelism and maximizing
hardware utilization.
% 
% data-intensive computer
% 
% Process-oriented programming naturally implements
% {\em object-based parallelism},
% % We show that process-oriented programming is
% % a powerful framework for parallel programming,
% and we propose it as a high-level replacement for multithreading, MPI and 
% many other languages, environments and tools currently used for
% parallel computations.
% %
% Potentially the most important impact of the process-oriented extension of 
% languages, such as C++ and Python, is a widespread adoption of parallel
% programming,
% as application developers realize the ability
% to easily create processes instead of using thread libraries
% and
% to place different
% objects on different CPU cores.
% 
% 

% We implemented a prototype process system using C++ with MPI
% and investigated process-oriented programming
% in the context of data-intensive computing.
% We have shown that a complex and efficient application can be built
% using only a few hundred lines of process-oriented code,
% which is
% equivalent to
% many thousands of lines of object-oriented code with MPI.

% We view a large data object as a collection of persistent processes.
% 
% The data-intensive Fourier transform
% computation was carried out on a small cluster
% of 8 nodes (96 cores).
\end{abstract}

% \newpage
\section{Introduction}
\label{sec:Introduction}

% From the early 1970s to approximately 2004 microprocessor
% manufacturers designed single processor CPUs,
% whose speed, clock frequency and energy density
% kept increasing from generation to generation.
% 
% By 2004 the clock frequencies of single processor CPUs
% 
% % For several decades 
% Beginning in the late 1960s
% the microprocessor industry produced 
% generation after generation of
% faster single-processor CPUs
% with increasing clock frequency and energy density.
% This process came to an end in 2004.
% 
% From the early 1970s 
% % microprocessor manufacturers produced 
% generation after generation of faster single-processor CPUs
% drove 

% The design of single processor CPUs
% Beginning in 2004 the microprocessor industry began producing
% multi-core CPUs

The first commercially available microprocessor CPUs appeared in the early
1970s.
These were single-processor devices that operated with a clock rate
of less than 1 MHz.
Over the course of the following three decades
increasingly
% more powerful,
faster and cheaper CPUs were built.
This was achieved in large part by
persistently
% relentlessly
increasing the clock rate.
By 2004 the CPU clock rates reached the 3-4 GHz range and heat dissipation
became a major problem.
In order to continue improving 
% application performance
the performance and the cost,
microprocessor designers turned to parallel computing.
Today,
nearly all computing devices (servers, tablets, phones, etc.)
are built using processors with multiple computing cores,
whose operating frequency is less than 3.5 GHz.
The industry move to
parallel computing
succeeded because practically every application
contains tasks that can be executed in parallel.
% there is lots of trivial parallelism that can be harvested.
And yet,
a decade later
the vast majority of computer programs are still being written 
for execution on a single processor,
and
parallel computation
% has been achieved
is being realized
primarily using threads,
sequential processes that share memory.

Parallel programming is 
% widely 
generally
recognized as difficult,
and has been a subject of extensive research
(see \cite{diaz2012survey},
\cite{herlihy2012art}, \cite{mattson2004patterns} and references therein).
The problem with threads is eloquently described in
\cite{lee2006problem}.
The author paints a bleak scenario:
% ``Intel, for example, has embarked on an active campaign 
% to get leading computer science academic programs 
% to put more emphasis on multithreaded programming. 
% If they succeed, and programmers make more intensive 
\begin{quote}
``If [...]
programmers make more intensive 
use of multithreading, the next generation of computers 
will become nearly unusable.''
\end{quote}
%
% The bulk of programming is sequential, and the bulk of non-sequential
% programming is threads.
%
In the scientific computing community
% parallel programming is typically done using
parallel programs are typically written in Fortran and C
with
OpenMP
\cite{openmp}
and MPI
% (Message Passing Interface)
\cite{mpi}.
% which
% have been developed
% since the early 1990s.
% as successful frameworks for
% shared memory and distributed memory programming.
Dozens of high-level languages for parallel programming have also been developed,
but presently none of them is widely used.
% Arguably, even implementation of embarrassingly parallel computations is not
% easy.
Even the so-called {\em embarrassingly parallel} computations are not
embarrassingly easy to implement.

% Parallel programming has been a subject of extensive research
% (\cite{herlihy2012art}, \cite{mattson2004patterns}),
% which led also to development
% % Extensive research in parallel programming led also to development
% of many specialized high-level languages,
% such as the PGAS programming languages
% \cite{yelick2007productivity}.
% But these languages are not widely used.

In this paper we develop a new framework for parallel programming,
which we call {\em process-oriented programming}.
It is based on the fundamental observation that any class
in an object-oriented language can be instantiated as a process,
accessible via a remote pointer
\cite{givelbergOOPP}.
% Such a process would act as a server
Such a process instantiates an object of the class and acts as a server
to other processes,
remotely
executing the class interface methods on this object.
The introduction of remote pointers
% is a powerful syntactic device
enables a straightforward extension of object-oriented
programming languages to process-oriented programming
with hardly any syntax additions.
We show that process-oriented programming is
an efficient framework for parallel programming,
and we propose it as a replacement for multithreading, MPI and 
% OpenMP.
many other languages, environments and tools currently used for
parallel computations.
We implemented a prototype system for running processes using C++ with MPI
and investigated process-oriented programming
in the context of data-intensive computing.

% We introduced processes into object-oriented programming in
% \cite{givelbergOOPP}
% by observing that every object can be instantiated as a process,
% which acts as a server
% which executes
% % executing
% the object's public interface methods.
% In this paper we
% % introduce 
% further develop 
% the concept of a process as a fundamental
% building block of 
% a program
% % an object-oriented program.
% %
% % In this paper we introduce processes into object-oriented
% % programming languages 
% and
% show that a complete framework for parallel programming
% with processes is obtained using only very small syntax extensions.
%
% We propose to extend object-oriented languages (C++, Python, Java, etc.)
% to include processes,
% and to adopt process-oriented programming as the principal model
% for parallel programming,
% replacing multithreading, MPI and OpenMP.
% % We expect the most valuable is:
% % process-oriented programming as a replacement for multithreading.
% % This is likely the most important use.
% % Design of operating systems!!!
% % We provide an extensive and detailed example of process-oriented
% % programming in a data-intensive application.
% In this paper 
% we restrict our investigation of process-oriented programming
% to data-intensive computing.
% % we investigate process-oriented programming in the context of
% % a data-intensive application.
% % We present a detailed example of the computation
% % of a large three-dimensional Fourier transform.

The rapid growth of generated and collected data 
in business and academia
% generated 
% by scientific simulations
% and high throughput measurement instruments
% requires performing 
creates demand for
increasingly complex and varied computations
with very large data sets.
The data sets are typically stored on hard drives,
% which are slow mechanical devices.
so the cost of accessing and moving 
small portions of the data set is high.
% Nevertheless, a large data set 
% can
% be stored on a large number
% of hard drives, and processed in parallel.
Nevertheless, 
a large number of hard drives can be used in parallel to
significantly reduce the amortized cost of data access.
%
% Therefore, the main difficulty in performing computations with large data
% sets lies in managing parallelism.
%
%
% For problems involving very large data sets the cost of moving the data
% is often the dominant cost.
% Previously the analysis of the complexity of computation focused on
% the number of arithmetic operations.
% The class of problems with 
% $O(N \log N)$ arithmetic complexity
% contains many important problems.
% For all practical values of $N$ this number is practical.
% It can be argued that these problems do not have many data movement
% operations.
In
\cite{givelberg2011architecture},
we argued that
{\em a data-intensive computer}
can be built, using
widely available (``commodity'') hardware components,
to solve general computational problems involving very large data sets.
% and programmed to efficiently solve general problems 
% of $O(N \log N)$ arithmetic complexity.

% The class of problems of 
% $O(N \log N)$ arithmetic complexity
% For many such problems the cost of data movement is higher than
% the cost of arithmetic operations.

% We considered the general problem of computation with very large
% data sets in
% \cite{givelberg2011architecture},
% and proposed that a data-intensive computer can be built using
% % standard 
% widely available (``commodity'') hardware components,
% and programmed to efficiently solve general problems of
% $O(N \log N)$ arithmetic complexity.
% For many such problems the cost of data movement is higher than
% the cost of arithmetic operations.
% and it is the main challenge addressed
% % by the research presented here.
% by this paper.
The primary challenge in the construction of the data-intens\-ive computer
lies in software engineering.
The software framework must balance programmer productivity
with efficient code execution,
i.e.
big data applications with complex data access patterns
must be realizable using a small amount of code,
and this code must attain high data throughput, using massive parallelism.
In this paper we aim to demonstrate that process-oriented programming
is the right framework for the realization of the data-intensive computer.
% \begin{itemize}
% \item
% The programmer must be able to create applications using
% a small amount of code.
% \item
% Problems requiring complex data access patterns should
% be efficiently computed with high data throughput realized
% by massive parallelism.
% \end{itemize}
% only a few hundred lines of code.
% enable the programmer to create
% applications using 
% high programmer productivity
% (the programmer can quickly generate complex code),
% while also enabling the programmer to generate efficient code.
% Which means that problems with complex data access patterns can
% be computed with high degree of parallelism, achieving
% very high data rates.

% In this paper we further develop the process framework 
% and demonstrate that it
% is suitable for the realization of the data-intensive computer.
%
% The programming framework for the data-intensive computer
% is object-oriented processes, which we introduced in
% \cite{givelbergOOPP}.
% In this paper we demonstrate an application of these ideas
% in data-intensive computing.
%
% While the Fourier transform is of great
% importance in many areas of science and engineering,
% for our study
% we chose the computation of a large 
% three-dimensional Fourier transform 
% as a prototype of a difficult data-intensive problem.
%
We chose the computation of a large 
three-dimensional Fourier transform 
as the subject of our study
primarily because it can be considered
as a prototype of a difficult data-intensive problem.
%
% Currently we do not have a compiler implementing processes,
% so we emulate the process framework 
% using approximately 15000 lines of C++ with MPI
% \cite{mpi}.
%
% Although we do not currently have a compiler supporting processes,
% we emulate the process framework using C++ and MPI
% \cite{mpi}.
%
% The large scale data-intensive computation is a test problem,
% and right now we emulate processes using C++ and MPI
% \cite{mpi}.
% We are implementing a compiler and a run time system.
%
We show that using processes
% Using 
% a compiler and a run-time system supporting processes,
our application 
% can be realized with 
% requires 
can be realized with
only a few hundred lines of code,
which are equivalent to
% instead of 
approximately 15000 lines of C++ with MPI.
% Using the process language extension the code would be only a few hundred
% lines.
% We demonstrate 
We also show 
that even with the complex data access patterns
required for the computation of the 3D Fourier transform,
our code attains very high data throughput
by achieving massive parallelism
and maximizing hardware utilization.

In section
\ref{sec:DataSet}
we describe a simple model of storage of a data set as a collection of
data pages on multiple hard-drives.
This model is used in the examples
in section
\ref{sec:Processes},
where we introduce processes.
The process-oriented implementation of the Fourier transform 
is described in section
\ref{sec:FourierTransform}
and the efficiency of the computation
is analyzed
in section
\ref{sec:Computation}.
We conclude with a discussion in section
\ref{sec:Conclusion}.

Our presentation uses C++, but can be easily applied to any object-oriented
language.
{\tt \color{blue} size\_t} is a large non-negative integer type used
in C++ to represent the size of a data object in bytes.

% \newpage

\section{
The Data Set
}
\label{sec:DataSet}

\subsection{
Data Pages
}
\label{sec:DataPages}
We represent an 
$N_1 \times N_2 \times N_3$
array of complex double precision numbers 
as a collection of 
$NP_1 \times NP_2 \times NP_3$
pages,
where each page is a small complex double precision array of size
$n_1 \times n_2 \times n_3$.
% We assume a $N_1 \times N_2 \times N_3$-point
% array is broken up into $NP_1 \times NP_2 \times NP_3$
% pages of size $n_1 \times n_2 \times n_3$.
%
First, we define a {\tt Page} class which stores {\tt n} bytes 
of unstructured data:
\begin{lstlisting}
class Page
{
public:
	Page(size_t n, unsigned char * data);
	~Page();
protected:
	size_t n;
	unsigned char * data;
};
\end{lstlisting}
The {\tt ArrayPage} class is derived from
the {\tt Page} class to handle
three-dimensional complex double-precision array blocks:
\begin{lstlisting}[escapechar=@]
class ArrayPage:
	public Page
{
public:
	ArrayPage(
		int n1, int n2, int n3,
		double * data
	);
	// constructor that allocates data:
	ArrayPage(int n1, int n2, int n3);
	void transpose13();
	void transpose23();
	@\vdots@
private:
	int n1, n2, n3;
}
\end{lstlisting}
The {\tt ArrayPage} class
may include functions for local computation with the array page data,
such as the transpose functions that are important in the computation
of the Fourier transform
(see section \ref{sec:FFTProcesses}).
Throughout this paper the arrays have equal dimensions
and the distinct variables {\tt n1, n2} and {\tt n3} are maintained only
for the clarity of exposition.

\subsection{
Storage devices
}
\label{sec:StorageDevices}

We store data pages on hard drives, using a single large file
for every available hard drive.
The following {\tt PageDevice} class controls the hard drive I/O.
\begin{lstlisting}
class PageDevice
{
public:
	PageDevice(
		string filename,
		size_t NumberOfPages,
		size_t PageSize
	);  
	~PageDevice();
	void write(Page * p, size_t PageIndex);
	void read(Page * p, size_t PageIndex);
protected:
	string filename;
	size_t NumberOfPages;
	size_t PageSize;
private:
	int file_descriptor;
};
\end{lstlisting}
The implementation of this class creates a file {\tt filename} of
{\tt NumberOfPages * PageSize} bytes.
Pages of data are stored in the {\tt PageDevice} object
using a {\tt PageIndex} address, where {\tt PageIndex}
ranges from {\tt 0} to {\tt NumberOfPages - 1}.
The {\tt write} method copies a data page of size {\tt PageSize}
to the location with an offset {\tt PageIndex * PageSize}
from the beginning of the file {\tt filename}.
Similarly, the {\tt read} method 
reads a page of data stored at a given integer address 
in the {\tt PageDevice}.
Linux direct I/O functions are used in the class implementation.
For example,
the Linux {\tt open} function, used with the {\tt O\_DIRECT} flag,
attempts to minimize cache effects of the I/O to and from the specified file.

For array pages we define the {\tt ArrayPageDevice},
which extends {\tt PageDevice}, as follows:
\begin{lstlisting}[escapechar=@]
class ArrayPageDevice:
	public PageDevice
{
public:
	ArrayPageDevice(
		string filename,
		size_t NumberOfPages,
		int nn1, int nn2, int nn3
	):
		n1(nn1), n2(nn2), n3(nn3),
		PageDevice(
			filename,
			NumberOfPages,
			2 * n1 * n2 * n3 * sizeof(double)
		)
	{}
	void write_transpose13(Page * p, size_t PageIndex);
	void read_transpose13(Page * p, size_t PageIndex);
	void write_transpose23(Page * p, size_t PageIndex);
	void read_transpose23(Page * p, size_t PageIndex);
	@\vdots@
private:
	int n1, int n2, int n3;
};
\end{lstlisting}
The implementation of the transpose methods is very simple.
For example:
\begin{lstlisting}[escapechar=@]
void ArrayPageDevice::
	read_transpose13(
		Page * p, size_t PageIndex
	)
{
	read(p, PageIndex);
	p->transpose13();
}
\end{lstlisting}
In addition to the transpose methods
the {\tt ArrayPageDevice} class may provides various other methods for computing
% The {\tt ArrayPageDevice} class may include methods for computing
with {\tt n1 $\times$ n2 $\times$ n3} blocks of complex double precision data.
Furthermore, in section
\ref{sec:caching}
we extend {\tt ArrayPageDevice} to include
% server-side 
caching.

\subsection{
Page map
}
\label{sec:PageMap}

Since we use multiple hard drives to store a single large array object,
% The {\tt server} collection defined above
% can be used to store multiple arrays. 
we introduce the {\tt PageMap} class
to specify the storage layout of a given array.
The {\tt PageMap}
translates
logical array page indices
into storage addresses.
A storage address consists of an id of the server, corresponding
to the hard drive where the data page is stored, and the page index,
indicating the address of the page on that hard drive.
\begin{lstlisting}[escapechar=@]
typedef 
	struct 
	{
		int server_id;
		size_t page_index;
	}
	address;

struct PageMap
{
	virtual address PageAddress(
		int i1, int i2, int i3
	);
};
\end{lstlisting}
% A storage address consists of a server id and a page index,
% and any process can read an array page from the {\tt server} collection,
% as follows:
% % every logical index $(i_1, i_2, i_3)$ of an array page
% % into a physical address, consisting of a server id and a page index,
% which is used by the corresponding {\tt ArrayPageDevice} server 
% to determine the address of the page on disk.
% The {\tt PageMap} of the array determines how pages
% are read and written
% (see section \ref{sec:FFTPageMap}).
% % Reading and writing pages is done 
% An {\tt fft3} process 
% % reads a page by executing the
% % {\tt read}
% executes a remote
% method of an appropriate {\tt server} process,
% % for example:
% as follows.
%
% \begin{lstlisting}[escapechar=@]
% PageMap page_map;
% 
% int server_id
	% = page_map(i1, i2, i3).server();
% size_t page_index 
	% = page_map(i1, i2, i3).index();
% 
% ArrayPage * p;
% server[server_id]->read(p, page_index);
% \end{lstlisting}

In most applications the {\tt PageMap} would be a simple,
on-the-fly computable function,
but it is also possible to implement it using an array of pre-computed values.

% \newpage

\section{
Processes
}
\label{sec:Processes}

In this section we introduce processes into object-oriented
programming languages 
and
show that a complete framework for parallel programming
with processes is obtained using only very small syntax extension.

\subsection{
Process creation, destruction and remote pointers
}
\label{sec:CreationDestruction}

Programming objects can be naturally interpreted as processes.
Upon creation,
such a process instantiates the object
% of the class
and proceeds to act as a server
to other processes,
remotely
executing the class interface methods on this object.
% 
% Programming objects can be naturally interpreted as processes.
% A process representing an object
% acts as a server to other processes,
% remotely executing its class methods.
For example, a program running on the computing node {\tt machine0}
can create a new {\tt PageDevice} process on {\tt machine1},
as follows:
\begin{lstlisting}[escapechar=@]
size_t number_of_pages = 1024;
size_t page_size = 32 * 1024 * 1024;
char * remote_machine = "machine1";

PageDevice * storage
	= new(remote_machine)
		PageDevice(
			"pagefile", 
			number_of_pages, 
			page_size
		);
\end{lstlisting}
% This process would act as a server to other processes,
% remotely executing its class methods.
It can then generate a page of data and store it
on the remote {\tt machine1} using the remote pointer
{\tt storage}:
\begin{lstlisting}[escapechar=@]
Page * page = GenerateDataPage();
size_t PageAddress = 17;
storage->write(page, PageAddress);
\end{lstlisting}
%
% When this code is executed on
% {\em \color{mygreen}machine 0},
% a {\tt PageDevice} object is created on the remote
% {\em \color{mygreen}machine 1}
% and a page of data is stored in it.
%
% Superficially, the above program differs from the standard C++ only in the
% extension of the operator {\tt new}.
% The new {\tt new} allocates objects on remote machines,
% using the address of the remote machine specified inside parentheses.
% This particular choice of syntax 
% is not important and is only used here to illustrate the new idea.
% No new syntax is needed to execute methods on remote objects.
%
The 
% {\tt storage} process
new {\tt PageDevice} process 
% created on the remote machine
on {\tt machine1}
acts as a server which listens on a communications
port, accepts commands from other processes,
executes them and sends results back to the clients.
% the parent process, acting as a client,
% and sends results back to the client.
The client-server protocol is generated by the compiler from the class
desccription.
Remote pointer dereferencing triggers
a sequence of events,
that includes several client-server communications, 
data transfer and execution
of code on both the client and the server machines.

Process semantics and remote pointers extend naturally to simple
objects, as shown in the following example:
\begin{lstlisting}[escapechar=@]
double * data 
	= new(remote_machine) double[1024];
data[7] = 3.1415;
double x = data[2];
\end{lstlisting}
When this code is executed on {\tt machine0},
a new process is created on {\tt remote\_machine}.
This process allocates 
a block of {\tt 1024} doubles 
and deploys a server that communicates with the parent client
running on {\tt machine0}.
The execution of {\tt data[7] = 3.1415;} requires communication
between the  
% \linebreak
client and the server, including sending the numbers
{\tt 7} and {\tt 3.1415} from the client to the server.
Similarly, the execution of the following command leads to
an assignment of the local variable {\tt x} with a copy of
the remote double {\tt data[2]} obtained over the network using
client-server communications.
We emphasize that code execution is sequential:
% the programmer must have the confidence that 
each instruction, and all
communications associated with it, is completed before the following
instruction is executed.

Finally, we remark that
the notion of the class destructor in C++ extends natually to process
objects:
destruction of a remote object causes termination of the
remote process and completion
of the correspoding client-server communications.

\subsection{
Parallel computation
}

The sequential programming model requires an execution of an
instruction to complete before the next instruction is executed.
We defined remote method execution to conform to this model,
and therefore
the calling process is kept idle until it is notified that the remote
method has completed.
In order to enable parallel computation we
% propose 
introduce
the {\tt \color{blue} start} keyword to indicate that
% a simple mechanism that enables
the calling process may proceed with the execution of the next
statement without waiting for the current statement to complete.

{\em Example:}
A shared memory computation is constructed
by providing 
access to the previously defined {\tt data} block 
to several computing
processes:
% , leading to an example of a shared memory implementation:
\begin{lstlisting}[escapechar=@]
const int N = 64;
class ComputingProcess;
ComputingProcess * process_group[N];

for (int i = 0;  i < N;  i ++)
	start process_group[i] = new(machine[i])
		ComputingProcess(data);
\end{lstlisting}
% Although the {\tt data} block is shared among the processes,
% the computation is sequential.
% In section \ref{sec:ParallelComputation} we show how 
% this computation can be parallelized.
%
The {\tt \color{blue} start} keyword can be used also 
for remote method and remote function calls,
as is shown below.
% (see \ref{}).

An array of remote pointers defines a group of processes.
It is easy to assign ids to the processes and to make
them aware of the other processes in the group.
This enables subsequent inter-process communication 
by remote method execution.
% Extending the example above, we add the {\tt SetProcessGroup}
% method to the class {\tt ComputingProcess}:
Extending the example above, we assume that
{\tt ComputingProcess} is derived from
the following
{\tt ProcessGroupMember}
class.
\begin{lstlisting}[escapechar=@]
class ProcessGroupMember
{
public:
	int ID() const
		{ return id; }
	int NumberOfProcesses() const
		{ return N; }
	ProcessGroupMember ** ProcessGroup() const
		{ return group; }
protected:
	virtual void SetProcessGroup(
		int my_id,
		int my_N,
		ProcessGroupMember ** my_group
	);
private:
	int id;
	int N;
	ProcessGroupMember ** group; 
};
\end{lstlisting}
% A straightforward implementation of the {\tt SetProcessGroup}
% is a shallow copy:
% \begin{lstlisting}[escapechar=@]
% void ProcessGroupMember::SetProcessGroup(
	% int my_id,
	% int my_N,
	% ProcessGroupMember ** my_group
% )
% {
	% id = my_id;
	% N = my_N;
	% group = my_group;
% }
% \end{lstlisting}
The parameter {\tt my\_group}
of the {\tt SetProcessGroup} method
is a remote pointer to an array of remote processes,
so a shallow copy implementation of {\tt SetProcessGroup} 
% will result in future communications overhead.
will result in redundant future communications.
% so future reference
% to its members will result in additional communications.
The following deep copy implementation of
{\tt SetProcessGroup},
which  copies the entire remote array of remote pointers
to a local array of remote pointers,
is preferable:
\begin{lstlisting}[escapechar=@]
void ProcessGroupMember::SetProcessGroup(
	int my_id,
	int my_N,
	ProcessGroupMember ** my_group
)
{
	id = my_id;
	N = my_N;
	group = new ProcessGroupMember * [N];

	for (int i = 0;  i < N;  i ++)
		// remote copy:
		group[i] = my_group[i]; 
}
\end{lstlisting}

After instantiating the processes in {\tt process\_group}
the master process can form a process group as follows:
\begin{lstlisting}[escapechar=@]
for (int id = 0;  id < N;  id ++)
	process_group[i]->SetProcessGroup(
		i, N, process_group
	);
\end{lstlisting}
% Since a group of processes is likely to access common objects,
% a synchronization mechanism is necessary.
% and, typically, barrier functions are useful.
% A compiler-supported barrier method for arrays of
% processes achieves this goal.
Alternatively, the master process can execute 
the {\tt SetProcessGroup}
calls in parallel,
but
this requires synchronizing the processes at the end.
% We use barrier functions to synchronize processes.
A standard way to do this is using barrier functions.
A process executing the barrier function call must wait
until all other processes in the group execute a barrier function
call before proceeding with the execution of the next statement:
\begin{lstlisting}[escapechar=@]
for (int id = 0;  id < N;  id ++)
	start process_group[i]->SetProcessGroup(
		i, N, process_group
	);

process_group->barrier();
\end{lstlisting}

% For example,
% the processes of {\tt process\_group} 
% can be synchronized with
% \\
% {\tt process\_group->barrier(); }.
% \\
% an explicit compiler-supported barrier method for arrays of objects
% may be useful.
% % In our 
% For example, the processes belonging to the {\tt fft} array
% can be synchronized with
% % \\
% {\tt fft->barrier();}

\subsection{
% Transient and persistent processes
% Persistent processes
Persistent objects and processes
}
\label{sec:PersistentProcesses}

We view a large data set as a collection of
{\em persistent processes},
which provide access to portions of the data set, 
as well as methods for computing with it.
Persistent processes are objects that can be destroyed only by
explicitly calling the destructor.
The runtime system is responsible for storing process representation,
and activating and de-activating processes, as needed.
Processes can be accessed
using a symbolic object address,
similar to addresses used by the Data Access Protocol (DAP)
\cite{gallagher2004data},
for example:
\begin{lstlisting}[escapechar=@]
PageDevice * page_device =
	"http://data/set/PageDevice/34";
\end{lstlisting}
% In section
% \ref{sec:TheArray}
% we describe an example of an {\tt Array} class,
% a complex class containing pointers to 
% We assume that all processes are persistent.
Because
persistence is especially important for large data objects,
in this paper we assume that all processes and all objects
are persistent.

% This new syntax is preferable to the more conventional
% way of providing the object address in the constructor:
% % A more conventional syntax is to provide the object address
% % in the constructor:
% \begin{lstlisting}[escapechar=@]
% PageDevice * page_device =
	% new PageDevice(
		% "http://data/set/PageDevice/34"
	% );
% \end{lstlisting}
% % trouble is people may want to use this signature for sth else
%
%
% We assume that all processes are persistent by default,
% except for transient processes, which are processes that have
% no public interface.
% When a derived class inherits a public interface, its processes
% will be persistent.
% A transient process does not need to be explicitly destroyed; it
% terminates after its constructor has been executed.
%%
% A transient processes is a process 
% % that receives a {\tt \color{blue} delete}
% % instruction when its parent process terminates.
% that has no public interface.

% The {\tt PageDevice} and {\tt ArrayPageDevice} classes give rise to persistent
% processes.
% OK: persistence is a property of the object, not a class.
% Therefore, some processes may be instantiated as persistent,
% while others as transient.
% It is easiest to assume that all processes are persistent by default.
% No need to introduce new keyword.

\section{
% Fourier transform computation
% Fourier transform processes
% Processes for Fourier transform 
A program to compute the
Fourier transform 
}
\label{sec:FourierTransform}

% We now demonstrate the process concepts with the FFT
% example.
% Fourier transform was chosen to demo data-intensive application.
% 
% We complete the description of the data set that 
% we started in \ref{}.
% The {\tt Array} class provides basic client services
% for accessing array data.
%

% 
\begin{figure}[t]
\includegraphics[width=\columnwidth]{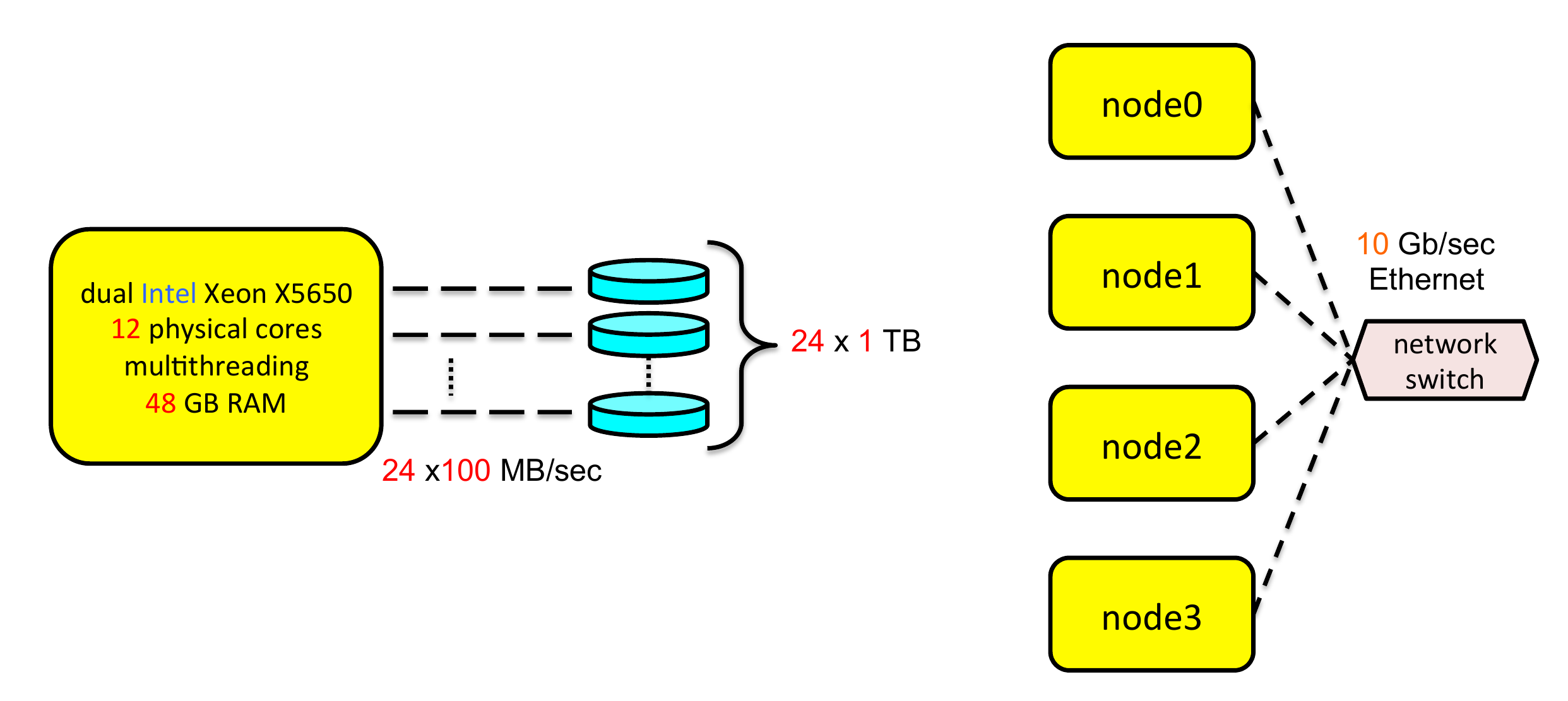}
\caption{
Cluster configuration for data intensive computing:
The nodes are interconnected by a high speed network (10 Gb/sec Ethernet).
Each node has a 12 core Intel Xeon processor with 48 GB of RAM
and 24 attached 1 TB hard-drives.
Each hard drive has an I/O throughput in the range of 100-150 MB/sec.
}
\label{fig:cluster}
\end{figure}

In this section we show how to use processes in a data-intensive
application on a cluster, such as the one shown in Figure
\ref{fig:cluster}.
We chose a large three-dimensional Fourier transform 
as an example of a data-intensive application
and we give an extensive description of the process-oriented
code.
% We show in detail an elaborate large-scale application,
% and how it can be programmed using the new process-oriented style
% of programming.
We consider the situation where
% In a typical situation 
the array data set is represented by a group
of processes, and the Fourier transform is a separate application
whose processes interact with the array processes.
% consisting of a different group of processes.
In section 
\ref{sec:caching}
we show how the efficiency of the parallel computation
can be improved using server-side caching.

% We assume a large array data set is created by by some application,
% so there is a set of persistent processes for it.
% Then we have another application that performs a Fourier transform
% on the array, so this application has its own processes.
% We first present {\tt Array} processes, then FFT processes.
% We then show how the parallel computation could be 
% % improved
% made more efficient
% by using server-side caching.

\subsection{
The Array
}
\label{sec:TheArray}

We described a method of storing a large data set
on multiple hard drives
in section
\ref{sec:DataSet}.
We now
% Here we
define the {\tt Array} class that is used
in the Fourier transform computation.
% We assume a $N_1 \times N_2 \times N_3$-point
% array is broken up into $NP_1 \times NP_2 \times NP_3$
% pages of size $n_1 \times n_2 \times n_3$.

The {\tt Array} class
describes a complex double precision three-dimensional array. 
% An {\tt Array} object 
% specifies the type of the array
% (i.e. its three-dimensional domain),
It specifies the array domain,
its storage layout and data access methods.
We first define an auxiliary
% {\tt Domain} 
class to describe rectangular 
three-dimensional domains:
% subdomains of the array domain:
\begin{lstlisting}[escapechar=@]
class Domain
{
public:
	Domain(
		int N11, int N12,
		int N21, int N22,
		int N31, int N32
	);
	@\vdots@
};
\end{lstlisting}

The collection of hard drives attached to the cluster nodes
can be turned into a distributed disk-based random access memory
by launching a {\tt PageDevice} process for every hard-drive
on the cluster node to which this hard-drive is attached,
but
% Furthermore, class inheritance
% makes it possible to define processes
% in terms of previously defined processes.
% naturally extends to processes.
for computations with arrays we launch {\tt ArrayPageDevice} processes
that enable us to perform some of the array computations
``close to the data'':
% ArrayPageDevice * server = new ArrayPageDevice * [N];
% const size_t DiskCapacity =
	% 960 * 1024 * 1024 * 1024;
% const size_t PageSize =
	% N1 * N2 * N3 * 2 * sizeof(double);
% const int NumberOfPages =
	% DiskCapacity / PageSize;
% const int NumberOfServers = 96;
\begin{lstlisting}[escapechar=@]
ArrayPageDevice ** page_server =
	new ArrayPageDevice * [NumberOfServers];

for (
	int i = 0;  i < NumberOfServers;  i ++
)
	page_server[i] = new(machine[i])
		ArrayPageDevice(
			filename[i],
			NumberOfPages,
			N1, N2, N3
		);
\end{lstlisting}
% Furthermore,
% launching a collection of {\tt ArrayPageDevice} processes on the cluster
% would provide additional local services for computation with array blocks.
% The {\tt ArrayPageDevice} processes could be launched either instead
% of {\tt PageDevice} processes, or alongside existing {\tt PageDevice} 
% processes.
The {\tt ArrayPageDevice} processes could be launched instead
of the {\tt PageDevice} processes, however in most applications it would
be advantageous to launch them alongside existing {\tt PageDevice} 
processes.
In this case {\tt ArrayPageDevice} must include a constructor
that takes a pointer to an existing {\tt PageDevice} process
as a parameter.
There would be essentially no communication overhead
when the {\tt ArrayPageDevice} process is launched on the same node
as the corresponding {\tt PageDevice} process.
The {\tt PageDevice::read} method, for example,
will copy a page from a hard drive 
directly into a memory buffer that is accessible 
by the 
% derived 
corresponding {\tt ArrayPageDevice} process.

% \begin{lstlisting}[escapechar=@]
% const int NumberOfServers = 96;
% const char * machine[NumberOfServers];
% const char * filename[NumberOfServers];
% 
% machine[0] = "node0";
	% filename[0] = "/24/disk0/pagefile";
% machine[1] = "node0";
	% filename[1] = "/24/disk1/pagefile";
	% @\vdots@
% machine[24] = "node1";
	% filename[24] = "/24/disk0/pagefile";
	% @\vdots@
% machine[95] = "node3";
	% filename[95] = "/24/disk23/pagefile";
% \end{lstlisting}

% The {\tt Array} class is a client class which
% enables
% a collection of processes
% to jointly construct the data set.
% It also provides basic data access and local computing services.
% The collection of the array data pages is handled by
% the {\tt page\_server} processes:
%
%
%
%
% In our Fourier transform computation
% we implemented the {\tt PageMap} as
% a small $128^3$-point array of 
% pre-computed values.
%
%
% \label{sec:FFTPageMap}
% For the Fourier transform
Array storage layout is determined by the {\tt PageMap} object.
The assignment of array pages to servers 
may affect
the degree of parallelism that can be achieved in the computation.
We use the circulant map,
which assigns array page
% $(i_1, i_2, i_3)$
{\tt (i1, i2, i3)}
to the server
{\tt (i1 + i2 + i3) \% NumberOfServers}.
% \[
% (i_1 + i_2 + i_3) \mod N_s.
% \]
% \begin{lstlisting}[escapechar=@]
	% (i1 + i2 + i3) % NumberOfServers.
% \end{lstlisting}
To complete the specification of the {\tt PageMap},
we assign the first available {\tt PageIndex} address within the 
target {\tt PageServer}.
% Any well-defined {\tt PageIndex} address of the page 
% within the {\tt PageServer}
% can be used.

%
The {\tt Array} class provides methods for a client process
to compute over a small array subdomain.
The client may use a small (e.g. 4 GB)
memory buffer to assemble the subdomain 
from array pages that reside within
{\tt page\_server} processes,
% {\tt ArrayPageDevice} processes, 
as determined by the array {\tt pagemap}.
% An {\tt Array} client uses a small (e.g. 4 GB)
% memory buffer to compute
% on a subdomain of the array data set.
% The subdomain is assembled from array pages that reside within
% {\tt page\_server} processes,
% % {\tt ArrayPageDevice} processes, 
% as determined by the array {\tt pagemap}.
%
	% public ProcessGroupMember
	% Array(char * ArrayFilename);
\begin{lstlisting}[escapechar=@]
class Array		// complex double 
{
public:
	Array(
		Domain * ArrayDomain, 
		Domain * PageDomain,
		int NumberOfServers, 
		ArrayPageDevice ** page_server,
		PageMap * pagemap
	);
	~Array();
	void read(Domain * d, double * buffer);
	void write(Domain * d, double * buffer);
	void read_transpose13(
		Domain * d, double * buffer
	);
	void write_transpose13(
		Domain * d, double * buffer
	);
	void read_transpose23(
		Domain * d, double * buffer
	);
	void write_transpose23(
		Domain * d, double * buffer
	);
private:
	Domain * ArrayDomain;
	Domain * PageDomain;
	int NumberOfServers; 
	ArrayPageDevice * page_server;
	PageMap * pagemap;
};
\end{lstlisting}
An {\tt Array} object is constructed by a single process,
which can then pass the object pointer to any group of processes.
Because an {\tt Array} is a persistent object, it can also be accessed
using a symbolic address,
as described in section
% Alternatively, the array can be accessed 
\ref{sec:PersistentProcesses}.

Transpose I/O methods are 
% useful for one-dimensional Fourier transform
% computations with long and narrow subdomains.
used to compute with long and narrow array subdomains
that are not aligned with the third dimension.
In section
\ref{sec:FFTProcesses}
we show an application of the transpose methods to
the computation of the Fourier transform.
% There are two ways to implement each of these methods.
% For example,
% % the implementation of
% {\tt read\_transpose13}
% can assemble the data by executing {\tt read\_transpose13}
% on appropriate {\tt page\_server} processes.
% Alternatively, the computation of the transpose can be done
% on the client side 
% by executing the server {\tt read} methods,
% followed by {\tt transpose13} on the received pages.
%
The transpose of array pages can be computed either on the client
or on the servers.
For example,
the implementation of
{\tt Array::read\_transpose13}
can assemble the data by executing
{\tt ArrayPageDevice::read\_transpose13}
on the appropriate {\tt page\_server} processes.
Alternatively,
it can execute {\tt ArrayPageDevice::read} methods on these servers,
followed by {\tt ArrayPage::transpose13} on the received pages.

\subsection{
FFT Processes
}
\label{sec:FFTProcesses}

We compute the three-dimensional Fourier transform
using three separate functions, 
{\tt fft1, fft2} and {\tt fft3},
% each computing 
each performing 
% a collection of 
one-dimensional Fourier transforms
along the corresponding dimension.
The functions {\tt fft1} and {\tt fft2} are similar
to {\tt fft3}, except that subdomains are read and written
using the I/O transpose operations of the {\tt Array} class.
Having read and transposed the data into a local memory buffer,
Fourier transforms are computed along the third dimension.
The result is then transposed back and written to the array.
% We therefore describe the {\tt fft3} function below.
We therefore restrict our description below to the {\tt fft3} function.

% Because our main goal here is to demonstrate a method 
% to carry out data-intensive computations,
% we made no attempt to optimize the Fourier transform
% algorithm with respect to the number of data movement operations.
% Instead, we chose the following simple algorithm.

The computation of the {\tt fft3} function is performed
in parallel using several {\tt FFT3} client processes:
% An {\tt FFT3} process is therefore a transient process which inherits from
% {\tt ProcessGroupMember}:
%
\begin{lstlisting}[escapechar=@]
class FFT3:
	public ProcessGroupMember
{
public:
	FFT3(int sign, Array * a);
	~FFT3() { delete buffer; }
	void ComputeTransform();
private:
	double * buffer;
	int sign;
	Array * a;
};
\end{lstlisting}
We divide the array into {\tt n}
slabs along the first dimension.
The master process launches
{\tt n FFT3} processes,
assigning each process to a slab.
\begin{lstlisting}[escapechar=@]
FFT3 ** fft = new FFT3 * [n];

for (int i = 0;  i < n;  i ++)
	fft[i] = new(node[i]) FFT3(sign, a);

for (int i = 0;  i < n;  i ++)
	fft[i]->SetProcessGroup(i, n, fft);

for (int i = 0;  i < n;  i ++)
	start fft[i]->ComputeTransform();
\end{lstlisting}
% Assuming that the array is divided into
% $N_1 \times N_2 \times N_3$ pages,
Each process maintains a buffer 
that can hold a page line,
% that can hold a line of pages,
where a page line
% a page line of the array
is a collection of $NP_3$ pages with page indices
\[
\left\{
(i_1, i_2, i_3): ~~~ i_3 = 0, 1, \ldots, NP_3 - 1
\right\}.
\]
% where $(i_1, i_2)$ are fixed.
% The total storage required for a page line is 4 GB.
% requiring a 4 GB buffer.
		% @{\em of the entire page line;}@
% so the required buffer size is 4 GB.
For complex double precision array of $128^3$ pages,
with each page consisting of $128^3$ points,
the page line buffer is 4 GB.
Each process computes:
\begin{lstlisting}[escapechar=@]
void FFT3::ComputeTransform()
{
	Domain * PageLine;
	for @{\em every}@ PageLine @{\em in the slab}@
	{
		a->read(PageLine, buffer);
		@{\em call FFTW}@(sign, buffer);
		a->write(PageLine, buffer);
		ProcessGroup()->barrier();
	}
}
\end{lstlisting}
An {\tt fft} process assembles a page line by reading
the pages from appropriate page servers.
For each page line
one FFTW
\cite{frigo2005design}
function call
% computes a set of $128^2$ one-dimensional complex 
% double Fourier transforms of size $16384$ each.
computes a set of 
% $128^2$
$n_1 \times n_2$
one-dimensional complex 
double Fourier transforms of size
% $16384$
$N_3$
each.
The result pages are then sent back to the page servers
and stored on hard drives.
Although the {\tt fft} processes do not communicate with each other,
they share a common network and a common pool of page servers.
The barrier synchronization at the end of each iteration
% improves the overall parallel performance of the computation.
is not strictly necessary.

\subsection{
% Parallel data access
Parallel execution; caching
}
\label{sec:caching}

Each page line {\tt read} and {\tt write} operation consists of
disk I/O and network transfer, with disk I/O being significantly
more time consuming.
% Part of the disk I/O can take place in parallel with the FFTW
% computation.
We implement caching on the page server in order to carry out
part of the disk I/O in parallel with the FFTW computation.
%
% When several {\tt Array} clients perform read/write operations concurrently,
% they share the network bandwidth and the processing bandwidth of the
% {\tt ArrayPageDevice} servers.
%
% % the page map is constructed to do data page access in parallel,
% % but this means we have to synchronize access to the servers.
% % (What is the alternative to synchronization?
% % It is better to have some mechanism that requires no
% % synchronization, but achieves high degree of parallelism)
% % 
% 
% There are several things that can be done in parallel.
% Reading (writing) to disk can be done in parallel with computation.
% The scheme described above leads to contentions between clients
% that may be are asking for service from the same server.
% Servers could be allocated to clients to avoid contentions.
% This requires synchronization, and it would require deriving
% a parallel device driver that does synchronization.
%
\begin{lstlisting}[escapechar=@]
class ArrayDevice:
	public ArrayPageDevice
{
public:
	ArrayPageDevice(
		string filename,
		int NumberOfPages,
		int n1, int n2, int n3,
		int Nc
	):
		ArrayPageDevice(
			filename,
			NumberOfPages,
			n1, n2, n3,
		),
		NumberOfCachePages(Nc)
	{}
	void ReadIntoCache(size_t page_index);
	void read(Page * p, size_t page_index);
	@\vdots@
private:
	int NumberOfCachePages;
	Page ** cache;
};
\end{lstlisting}
An {\tt ArrayDevice} server is configured
with a cache that can hold a line of pages.
	% void DeleteFromCache(size_t page_index);
%
% It becomes clear now that the {\tt ReadIntoCache} method
% must be executed in parallel, so we need a piece of new syntax.
% \begin{lstlisting}[escapechar=@]
% ArrayDevice * d;
% size_t page_index = 33;
% start d->ReadIntoCache(page_index);
	% @\vdots@
% \end{lstlisting}
% The new keyword {\tt \color{blue} start} has the effect that method execution is
% invoked on the remote object, but the calling process is not
% waiting for its completion and proceeds with its computation.
% We introduce interesting semantics: no method can be executed on
% an object until a given method completes.
% I.e. if we are going to read the page from the server's cache,
% it will happen only after the server completed reading it into the cache.
% This way we know when the parallel method has finished.
%
%
% \begin{lstlisting}[escapechar=@]
% for (int i1 = 0;  i1 < n1;  i1 ++)
% for (int i2 = 0;  i2 < n2;  i2 ++)
% for (int i3 = 0;  i3 < n3;  i3 ++)
% {
	% int id = 
		% page_map(i1, i2, i3).server_id();
	% size_t i = 
		% page_map(i1, i2, i3).index();
% 
	% start server[id]->ReadIntoCache(i);
% }
% \end{lstlisting}
%
In order to use {\tt ArrayDevice} instead of {\tt ArrayPageDevice}
we add the following
{\tt ReadIntoServerCache} method to the {\tt Array} class:
\begin{lstlisting}[escapechar=@]
void Array::ReadIntoServerCache(
	Domain * domain
)
{
	for @{\em every page}@ (i1, i2, i3) @{\em in}@ domain
	{
		address a = 
			pagemap->PageAddress(i1, i2, i3);
		int id = a.server_id;
		size_t i = a.page_index;
		start server[id]->ReadIntoCache(i);
	}
}
\end{lstlisting}
The {\tt ArrayDevice:read} method
executes {\tt ArrayPageDevice::read} if the requested page
is not found in its cache; otherwise it returns (sends over the network)
the cached page.

The transform computation is now reorganized, so that
instructions to read the next line of pages are sent to the page
servers before the FFT of the current page line is started.
% The cache class now means that we can tell it to read
% into cache or just to read, and if it is in the cache
% then we only have network transfer time.
% When the cache gets full, it will be written over by
% the new things we read.
% 
% We can now restructure our computation as follows:
	% Domain * PageLine = @{\em first page line}@;
	% a->ReadIntoServerCache(PageLine);
\begin{lstlisting}[escapechar=@]
void FFT3::ComputeTransform()
{
	Domain * PageLine;
	for @{\em every}@ PageLine @{\em in the slab}@
	{
		a->read(PageLine, buffer);
		Domain * NextPageLine = @{\em next page line}@;
		start a->ReadIntoServerCache(NextPageLine);
		@{\em call FFTW}@(sign, buffer);
		a->write(PageLine, buffer);
		ProcessGroup()->barrier();
	}
}
\end{lstlisting}
After the completion of the FFT the servers are instructed
to write pages to hard drives.
An execution of the {\tt write} method will take place only after
the server has completed {\tt ArrayDevice::ReadIntoCache}.
The efficiency of server-side caching depends on the relative
timing of the FFT computation and the page I/O.
If necessary,
% {\tt ArrayDevice::write} 
{\tt write} 
and 
% {\tt ArrayDevice::Write\-FromCache}
{\tt Write\-FromCache}
methods
can be implemented in the {\tt ArrayDevice} class analogously.

\section{
Computation of the Fourier Transform
}
\label{sec:Computation}

% We computed a large data-intensive Fourier Transform
% on a cluster. We implemented processes in our software
% and we measured the performance of the system during 
% the computation.
In this section we describe the computation of a large (64 TB)
data-intensive Fourier transform on a small (8 nodes, 96 cores)
cluster.
We describe our prototype implementation of the process framework
and analyze the performance of the Fourier transform computation.

\subsection{
Software Implementation of Processes
}
\label{sec:SoftwareImplementation}

The Fourier transform application was developed
by completing 
the process-oriented code which is sketched in sections
\ref{sec:DataSet}
and
\ref{sec:FourierTransform},
translating it
into C++ with MPI and linking it against
an auxiliary library of tools for implementation of basic process functionality.
% a library of auxiliary tools for implementation of basic process functionality.
We describe the procedure for {\tt ArrayDevice} processes.
In order to instantiate 
{\tt ArrayDevice}
processes, we created an
{\tt ArrayDevice\_server} class
and a C++ program file 
{\tt ArrayDevice\_process.cpp}
containing the following code:
\begin{lstlisting}[escapechar=@]
// start the server and disconnect from parent process
int main(int argc, char **argv)
{
	MPI_Init(&argc, &argv);
	MPI_Comm parent;
	MPI_Comm_get_parent(&parent);
	ArrayDevice_server * s =
		new ArrayDevice_server();
	MPI_Comm_disconnect(&parent);
	MPI_Finalize();
}
\end{lstlisting}
In addition, we implemented the function
\begin{lstlisting}[escapechar=@]
void LaunchProcess(
    const string & process_file_name,
    const string & machine_address,
    string & server_address
);
\end{lstlisting}
{\tt LaunchProcess} uses 
{\tt MPI\_Comm\_spawn} to create an MPI process by running the executable
{\tt process\_file\_name} on the remote machine specified by
{\tt machine\_address}.
The launched process starts a server which uses {\tt MPI\_Open\_port}
to open a port and return the port address in the {\tt server\_address}
output parameter.
The server process then disconnects from the launching process.

In order to implement remote method execution
we also created an
{\tt ArrayDevice\_client} class.
The {\tt ArrayDevice\_client} and the {\tt ArrayDevice\_server}
classes are automatically constructed from the
\\
{\tt ArrayDevice}
class.
For the purposes of this example we use a simple name mangling scheme
to first generate the following {\tt ArrayDevice\_interface} class:
	% // meta-description of ArrayDevice
\begin{lstlisting}[escapechar=@]
class ArrayDevice_interface
{
	@\vdots@
public:
	virtual void ArrayDevice_create(		// constructor
		string filename,
		size_t numberofpages,
		size_t pagesize
	);
	virtual void ArrayDevice_destroy();		// destructor
	virtual void ArrayDevice_write(Page * p, size_t page_index);
	virtual void ArrayDevice_ReadIntoCache(
		int n,
		size_t * page_index
	);

	// commands used in client-server protocol
	enum ArrayDevice_command
	{
		create_CMD = 0,
		destroy_CMD = 1,
		write_CMD = 2,
		ReadIntoCache_CMD = 3
	};
	@\vdots@
};
\end{lstlisting}
The {\tt ArrayDevice\_interface} class contains meta-information
obtained from the {\tt ArrayDevice} class.
% This information can be obtained automatically from a parser.
%
% For the purposes of this example we use a simple name mangling scheme
% to generate {\tt ArrayDevice\_client} methods that are constructed from
% the public methods of the {\tt ArrayDevice} class.
% The construction of the {\tt ArrayDevice\_server} class is completely
% analogous:
% These classes are derived from {\tt ProcessClient} and {\tt ProcessServer}
% which implement general protocols for client-server communications.
Both the {\tt ArrayDevice\_client} and the {\tt ArrayDevice\_server}
classes
are derived from {\tt ArrayDevice\_interface}.
% \\
{\tt ArrayDevice\_client} is also derived from a general
{\tt ProcessClient} class, and similarly {\tt ArrayDevice\_server}
is derived from {\tt ProcessServer}.
Jointly, the {\tt ArrayDevice\_client}, {\tt ArrayDevice\_server} pair
implement remote procedure calls (RPC) for {\tt ArrayDevice}.
% The client-server communication protocol is implemented in
% {\tt ProcessClient} and {\tt ProcessServer}, and it uses
% the meta-description of the {\tt ArrayDevice} class that is provided
% in {\tt ArrayDevice\_interface}.
The client-server communications protocol for {\tt ArrayDevice}
uses the 
% \\
{\tt ArrayDevice\_interface}
class and the general-purpose functionality implemented 
in {\tt ProcessClient} and {\tt ProcessServer}.
The client-server implementation uses a simple object serialization
library to send method parameters and results over the network.
This serialization library is also used to implement a rudimentary
file-based object persistence mechanism.
% , with serialized objects being written to
% and read from files.
Information is sent over the network using an MPI-based communications
software layer.

% The implementation of processes 
% % for the computation of the Fourier transform
% for the Fourier transform application
% requires construction of 
% only a handful of classes,
% % Only a handful of classes must be constructed for
% % the implementation of processes
% % for the Fourier transform application,
% but the procedure outlined above can be 
% incorporated into a compiler
% % and applied to
% to build
% more complex applications.
% In our example
% the {\tt ArrayDevice\_interface} class would be
% constructed using information obtained from the parser,
% and the auxiliary classes
% % and function calls 
% would be generated automatically.
% % the meta-description of the {\tt ArrayDevice} class can be obtained from
% % the parser and the auxiliary classes and function calls can be generated
% % automatically.
% %
The translation procedure we implemented for the processes of the Fourier transform
application can be extended and incorporated into a C++ compiler.
%
% The translation process outlined above 
It
converts several hundred lines
of process-oriented code into
a C++/MPI application for computing Fourier transform,
which is approximately 15000 lines long.
% approximately about 15000 lines of C++/MPI code,
% which comprise the application for computing Fourier transform
% The software 
% was implemented in
% C++ with Intel MPI.
A very small subset of MPI functions is used.
The following is the almost complete list:
\begin{itemize}
\item
{\tt
MPI\_Send, MPI\_Recv
}
-- for inter-process communication,
\item
{\tt
MPI\_barrier
}
-- for process synchronization,
\item
{\tt
MPI\_Open\_port, MPI\_Close\_port, MPI\_Comm\_accept
}
-- to implement client-server functionality,
\item
{\tt
MPI\_Comm\_spawn, MPI\_Info\_create, MPI\_Info\_set
}
-- to spawn processes on specified target machines,
\item
{\tt
MPI\_Comm\_connect, MPI\_Comm\_disconnect,
MPI\_Comm\_get\_parent
}
-- to manage process connections.
% MPI\_Init, MPI\_Finalize,
\end{itemize}
Both MPICH and Intel MPI were used in our computation,
and with both implementations
we encountered problems with some MPI functions.
The most significant bugs were found in
 {\tt MPI\_Comm\_spawn}
and {\tt MPI\_Comm\_disconnect}.

% We emulated processes in C++ with MPI.
% If we had a compiler this could be done with only a few hunderds of lines
% of code.

% I think the compiler will have to do a *lot* of work. 
% I don't understand how the linking can work, 
% how it can be distributed amongst the machines to be used, 
% which can and will be specified at run time. 
% There are significant issues around run time environments 
% (shared libraries etc.) that need to be solved. 
% Is this truly distributed, or just for multi-core machines? 
% If fully distributed, how are potentially large latencies handled? 
% How are the communications between the processors implemented 
% - some sort of shared memory space, TCP/IP, exotic fabrics? 
% Can I use my existing cluster to run these applications? 
% What happens if one of the machines breaks during execution of the parallel app? 
% How does one debug a thing like this? 
% Who provides the management software that shows 
% a user the state of execution on all the participating machines?

% My feeling is that the abstraction is elegant at the code level, 
% but something still has to do the heavy lifting that the CS Parallel 
% Programming folks work on - there are no short cuts, I'm afraid. 
% (I also think you will piss off that community if you claim you have 
% rendered their profession irrelevant, which is probably not 
% what you want to do if you're seeking their support!)

% \subsection{The Hardware}
% \label{sec:Hardware}

\subsection{
Fourier Transform Computation
}
\label{sec:Results}

We used the process 
% framework 
prototype implementation
to carry out a computation of
% We carried out a computation of
% We describe a computation of
% We use the process-oriented code to compute 
the Fourier transform of
a $(16K)^3$-point array
of complex double precision numbers.
(We use the notation $1K = 1024, \, 16 K = 16384 = 128 \times 128$.)
The total size of the array is 64 TB.
%
% \begin{description}
% \item[Hardware]
The computations were carried out on a cluster of 8 nodes,
interconnected with a 10 Gb/sec Ethernet,
similar to the cluster depicted in Figure \ref{fig:cluster}.
Each computational node has an
Intel\textsuperscript{\textregistered}
Xeon\textsuperscript{\textregistered}
X5650  12 core CPU with hyperthreading,
rated at 2.67GHz,
and 24 attached 1 TB hard drives.
The hard drives are manufactured by Samsung, model SpinPoint F3 HD103SJ,
% Model Family:     SAMSUNG SpinPoint F3
% Device Model:     SAMSUNG HD103SJ
with manufacturer listed latency of 4.14 ms
and average seek time of 8.9 ms.
% The 1 TB drive comes with a 32 MB buffer memory.
Benchamark hard drive read and write throughput 
is reported at over 100 MB/sec.
The $(16 \, K)^3$-point array was broken up 
% We break up an array of $(16 \, K)^3$ points
% ($16 \, K = 16384 = 128 \times 128$)
into 
% The array is therefore stored as a collection of 
$128^3$
{\tt ArrayPage}
% objects
pages 
of $128 \times 128 \times 128$ points each.
The resulting page size is 32 MB.
The choice of the page size is constrained by the latency 
and seek time of the system's hard drives:
the smaller the page size, the lower the overall disk I/O throughput.
We measured a
typical page read/write time 
% on our system is 
in the range
of 0.25-0.35 sec.

We used 4 of the 8 available nodes to store the {\tt Array} object,
creating 24 {\tt ArrayDevice} processes on each node,
one process for each available hard drive.
Because {\tt ArrayDevice} processes are primarily dedicated to disk I/O
it is possible to run 24 processes on a 12-core node.
The process framework makes it easy to shift computation closer to data
by extending the {\tt ArrayDevice} class, and in such case relatively
more powerful CPUs may be needed to run the server processes.

% Our storage layout for the {\tt Array} object uses 4 of the 8 available
% nodes, with 24 {\tt ArrayDevice} processes on each node,
% one process for each available hard drive.
% The other 4 nodes were used
We used the other 4 nodes
to run 16 processes of the Fourier transform application,
4 processes per node.
Each of the 16 Fourier transform processes was assigned an array slab
of $8 \times 128 \times 128$ pages.
The process computes the transform of its slab line by line in
$8 \times 128 = 1024$
iterations.
Although the 16 processes are independent of each other,
they compete among themselves for service from 96 page servers.

% We measured the wall clock duration of {\tt ComputeTransform} iterations.
The wall clock time for a single iteration of {\tt ComputeTransform}
(see section \ref{sec:caching})
generally ranged from 68 to 78 seconds,
with the average of approximately 73 seconds.
The total speed of data processing (including reading, computing and writing
the data) has been therefore close to 1 GB/sec.
In the next section we analyze the performance in detail and indicate
a number of ways to substantially improve it.

\subsection{
Performance analysis
}
\label{sec:PerformanceAnalysis}

The computation of the Fourier transform was completed in the course
of several long (over 10 hours) continuous runs.
Our implementation of persistence mechanism for processes
made it possible to stop and restart the computation multiple times.
The primary reason for long runs was to test the stability and robustness
of our implementation.
We instrumented the code
to measure the utilization of the system's
components: the network, the hard drives and the processors.
Because
the results did not vary substantially over the course of the computation,
we present a detailed analysis of a typical iteration of {\tt ComputeTransform}.

The synchronization of the processes at the end of each iteration
is not strictly necessary, but we found that it did not significantly 
affect performance, and made the code easier to analyze.
On the other hand, we found that introducing additional barriers
within the iteration would slow down the computation significantly.
We now present detailed measurements of the component phases of the iteration.

% Our primary goal was to demonstrate that the process framework
% generates efficient code, but this code can be optimized much
% further and we also do not present scalability study.
% We do however make comments about what would be a better 
% hardware configuration.
% Detailed measurements of the basic parts of a typical iteration
% are presented in
% Figures
% \ref{fig:PageLineReadTime}
% --
% \ref{fig:writeTime}.

% \begingroup
    % \fontsize{10pt}{12pt}\selectfont
    % \begin{verbatim}  
        % % how to set font size here to 10 px ?  
    % \end{verbatim}  
% \endgroup

% \makeatletter
% \newcommand{\verbatimfont}[1]{\def\verbatim@font{#1}}%
% \makeatother

At the beginning of every iteration
each process reads a page line
consisting of 128 pages,
which are evenly distributed 
among the 96 servers by the circulant map
(see section \ref{sec:TheArray}).
Except for the first iteration, the required pages
have already been read from the hard drive and placed
% are already located 
in the memory of the corresponding servers.
Each server has
either 21 or 22
pages to send to the clients.
In total, 64 GB of data is sent from the servers to the clients.
Accordingly, the combined size of the server caches in each of the server nodes
is slightly more than 16 GB.
We timed the parallel reading of page lines by
% Timing measurements of the parallel reading of page lines by 
the 16 client processes in a typical time step:
%
% \begin{figure}
\begin{lstlisting}[escapechar=@,basicstyle=\scriptsize\ttfamily,frame=l]
15: Array::read 11520-11647 x 2432-2559 x 0-16383:: 4 GB, 17.3638 sec, 235.894 MB/sec
12: Array::read 9216-9343 x 2432-2559 x 0-16383:: 4 GB, 18.6695 sec, 219.395 MB/sec
0: Array::read 0-127 x 2432-2559 x 0-16383:: 4 GB, 21.6104 sec, 189.538 MB/sec
1: Array::read 768-895 x 2432-2559 x 0-16383:: 4 GB, 22.2821 sec, 183.824 MB/sec
11: Array::read 8448-8575 x 2432-2559 x 0-16383:: 4 GB, 22.4682 sec, 182.302 MB/sec
8: Array::read 6144-6271 x 2432-2559 x 0-16383:: 4 GB, 22.7035 sec, 180.412 MB/sec
14: Array::read 10752-10879 x 2432-2559 x 0-16383:: 4 GB, 22.9213 sec, 178.698 MB/sec
5: Array::read 3840-3967 x 2432-2559 x 0-16383:: 4 GB, 27.1039 sec, 151.122 MB/sec
6: Array::read 4608-4735 x 2432-2559 x 0-16383:: 4 GB, 27.0914 sec, 151.192 MB/sec
2: Array::read 1536-1663 x 2432-2559 x 0-16383:: 4 GB, 27.2604 sec, 150.255 MB/sec
4: Array::read 3072-3199 x 2432-2559 x 0-16383:: 4 GB, 28.1986 sec, 145.256 MB/sec
3: Array::read 2304-2431 x 2432-2559 x 0-16383:: 4 GB, 28.3351 sec, 144.556 MB/sec
7: Array::read 5376-5503 x 2432-2559 x 0-16383:: 4 GB, 28.9623 sec, 141.425 MB/sec
10: Array::read 7680-7807 x 2432-2559 x 0-16383:: 4 GB, 29.027 sec, 141.11 MB/sec
9: Array::read 6912-7039 x 2432-2559 x 0-16383:: 4 GB, 29.5628 sec, 138.553 MB/sec
13: Array::read 9984-10111 x 2432-2559 x 0-16383:: 4 GB, 30.1746 sec, 135.743 MB/sec
\end{lstlisting}
% \caption{
% Timing measurements of the parallel reading of page lines by 
% the 16 client processes in a typical time step.
% The first number in each row is the client process id,
% followed by the domain, domain size, the time it took to read it
% and the corresponding throughput.
% }
% \label{fig:PageLineReadTime}
% \end{figure}
The first number in each row is the client process id,
followed by the description of the domain, domain size, the time it took to read it
and the corresponding throughput.
%
% Figure
% \ref{fig:PageLineReadTime}
% shows the timing of this phase, measured by each client process
% during a typical time step.
The fastest process completes the execution of the {\tt Array::read}
method (including sending the command and the parameter to page servers)
in approximately 17.5 seconds, the slowest -- in about 30 seconds.
The faster processes proceed to start the execution
of {\tt Array::ReadIntoServerCache}
immediately after completion of {\tt Array::read}.
The typical aggregate throughput during the parallel execution of
{\tt Array::read} is therefore significantly larger than 2 GB/sec.
The maximal possible throughput to the four nodes computing the transform
is approximately 4 GB/sec.

In the {\tt Array::ReadIntoServerCache}
phase of the computation 16 clients send small messages to
96 page servers with instructions to read a total of $16 \times 128$ pages.
These commands are queued for execution in the servers,
so that the clients do not wait for the completion of the execution.
The time measurements of the parallel execution of
{\tt {\color {blue} start} Array::ReadIntoServerCache}
by the 16 client processes in a typical time step were:
% 5: >>> PageFile_object::PageFile_read_into_cache server id = 5 connected:: 0.01546 sec
% 13: >>> PageFile_object::PageFile_read_into_cache server id = 13 connected:: 0.014056 sec
% 14: >>> PageFile_object::PageFile_read_into_cache server id = 14 connected:: 0.012255 sec
%
%
% \begin{figure}
\begin{lstlisting}[escapechar=@,basicstyle=\scriptsize\ttfamily,frame=l]
14: Array::ReadIntoServerCache 10752-10879 x 2560-2687 x 0-16383:: 9.36148 sec
13: Array::ReadIntoServerCache 9984-10111 x 2560-2687 x 0-16383:: 2.10771 sec
10: Array::ReadIntoServerCache 7680-7807 x 2560-2687 x 0-16383:: 3.25717 sec
12: Array::ReadIntoServerCache 9216-9343 x 2560-2687 x 0-16383:: 9.5796 sec
8: Array::ReadIntoServerCache 6144-6271 x 2560-2687 x 0-16383:: 9.81484 sec
11: Array::ReadIntoServerCache 8448-8575 x 2560-2687 x 0-16383:: 13.6131 sec
9: Array::ReadIntoServerCache 6912-7039 x 2560-2687 x 0-16383:: 2.67452 sec
5: Array::ReadIntoServerCache 3840-3967 x 2560-2687 x 0-16383:: 5.16807 sec
2: Array::ReadIntoServerCache 1536-1663 x 2560-2687 x 0-16383:: 5.0223 sec
4: Array::ReadIntoServerCache 3072-3199 x 2560-2687 x 0-16383:: 4.08604 sec
0: Array::ReadIntoServerCache 0-127 x 2560-2687 x 0-16383:: 10.6741 sec
3: Array::ReadIntoServerCache 2304-2431 x 2560-2687 x 0-16383:: 3.94899 sec
1: Array::ReadIntoServerCache 768-895 x 2560-2687 x 0-16383:: 9.98827 sec
6: Array::ReadIntoServerCache 4608-4735 x 2560-2687 x 0-16383:: 5.16787 sec
15: Array::ReadIntoServerCache 11520-11647 x 2560-2687 x 0-16383:: 14.9185 sec
7: Array::ReadIntoServerCache 5376-5503 x 2560-2687 x 0-16383:: 3.27696 sec
\end{lstlisting}
% \caption{
% Time measurements of the parallel execution of
% {\tt {\color {blue} start} Array::ReadIntoServerCache}
% by the 16 client processes in a typical time step.
% }
% \label{fig:ReadIntoServerCacheTime}
% \end{figure}
%
% The measurements for the execution of 
% {\tt {\color{blue} start} Array::ReadIntoServerCache} is shown in
% Figure \ref{fig:ReadIntoServerCacheTime}.
%
% The timiings in Figure \ref{fig:ReadIntoServerCacheTime} are bad
These results are significantly worse than expected.
Our implementation of processes repeatedly establishes
and breaks client-server connections.
We found the {\tt MPI\_Comm\_connect} function to be very fast,
but with increasing number of client-server connections,
it sporadically
% occasionally 
performed hundreds of times slower than usual.
An implementation of caching connections is likely to reduce the total time
for this phase of the computation to about 3 seconds.

We timed the execution of the FFTW function call by every processor.
% \begin{figure}
\begin{lstlisting}[escapechar=@,basicstyle=\scriptsize\ttfamily,frame=l]
14: fftw 10752-10879 x 2432-2559 x 0-16383   5.53611 sec
0: fftw 0-127 x 2432-2559 x 0-16383   5.86336 sec
6: fftw 4608-4735 x 2432-2559 x 0-16383   5.91158 sec
4: fftw 3072-3199 x 2432-2559 x 0-16383   5.99485 sec
8: fftw 6144-6271 x 2432-2559 x 0-16383   6.01364 sec
12: fftw 9216-9343 x 2432-2559 x 0-16383   6.03876 sec
10: fftw 7680-7807 x 2432-2559 x 0-16383   6.06033 sec
2: fftw 1536-1663 x 2432-2559 x 0-16383   6.07081 sec
11: fftw 8448-8575 x 2432-2559 x 0-16383   6.38444 sec
7: fftw 5376-5503 x 2432-2559 x 0-16383   6.38421 sec
15: fftw 11520-11647 x 2432-2559 x 0-16383   6.42032 sec
3: fftw 2304-2431 x 2432-2559 x 0-16383   6.42975 sec
9: fftw 6912-7039 x 2432-2559 x 0-16383   14.001 sec
13: fftw 9984-10111 x 2432-2559 x 0-16383   14.5631 sec
5: fftw 3840-3967 x 2432-2559 x 0-16383   14.5582 sec
1: fftw 768-895 x 2432-2559 x 0-16383   14.878 sec
\end{lstlisting}
% \caption{
% Time measurements of the parallel execution of
% the FFTW library calls
% by the 16 client processes in a typical time step.
% }
% \label{fig:FFTTime}
% \end{figure}
%
The last 4 processes (9,13,5 and 1) ran on the same node.
Throughout our computation these 4 processes executed the FFTW
library function call significantly slower than the processes
running on other nodes.
Additional investigation of the configuration of this node
is needed to speed up the computation.

The reading of the pages on the server side is done concurrently
with the FFTW computation.
% The servers are completely independent of each other.
We include the measurements for a few of the 96 servers:
%
% 15: ArrayDevice::ReadIntoCache: 22 pages, 704 MB, 5.66957 sec, 124.172 MB/sec
% 55: ArrayDevice::ReadIntoCache: 21 pages, 672 MB, 5.62723 sec, 119.419 MB/sec
% 58: ArrayDevice::ReadIntoCache: 22 pages, 704 MB, 5.62373 sec, 125.184 MB/sec
% 13: ArrayDevice::ReadIntoCache: 21 pages, 672 MB, 5.58151 sec, 120.398 MB/sec
% 16: ArrayDevice::ReadIntoCache: 22 pages, 704 MB, 5.69196 sec, 123.683 MB/sec
% 10: ArrayDevice::ReadIntoCache: 22 pages, 704 MB, 5.59965 sec, 125.722 MB/sec
% 17: ArrayDevice::ReadIntoCache: 21 pages, 672 MB, 5.70713 sec, 117.747 MB/sec
% 8: ArrayDevice::ReadIntoCache: 21 pages, 672 MB, 5.60408 sec, 119.913 MB/sec
% 28: ArrayDevice::ReadIntoCache: 22 pages, 704 MB, 5.70762 sec, 123.344 MB/sec
% 1: ArrayDevice::ReadIntoCache: 21 pages, 672 MB, 5.62662 sec, 119.432 MB/sec
% \begin{figure}
% % \begingroup
    % % % \fontsize{8pt}{10pt}\selectfont
    % % \fontsize{6pt}{8pt}\selectfont
% % \begin{verbatim}
% % \begin{lstlisting}[escapechar=@,basicstyle=\tiny\ttfamily]
% % \begin{lstlisting}[escapechar=@,basicstyle=\scriptsize\ttfamily]
\begin{lstlisting}[escapechar=@,basicstyle=\scriptsize\ttfamily,frame=l]
@\vdots@
34: ArrayDevice::ReadIntoCache: 22 pages, 704 MB, 5.59864 sec, 125.745 MB/sec
25: ArrayDevice::ReadIntoCache: 21 pages, 672 MB, 5.614 sec, 119.701 MB/sec
57: ArrayDevice::ReadIntoCache: 22 pages, 704 MB, 5.58319 sec, 126.093 MB/sec
52: ArrayDevice::ReadIntoCache: 22 pages, 704 MB, 5.59719 sec, 125.777 MB/sec
40: ArrayDevice::ReadIntoCache: 22 pages, 704 MB, 5.61367 sec, 125.408 MB/sec
73: ArrayDevice::ReadIntoCache: 21 pages, 672 MB, 5.57943 sec, 120.442 MB/sec
21: ArrayDevice::ReadIntoCache: 22 pages, 704 MB, 5.64487 sec, 124.715 MB/sec
33: ArrayDevice::ReadIntoCache: 22 pages, 704 MB, 5.63134 sec, 125.015 MB/sec
89: ArrayDevice::ReadIntoCache: 21 pages, 672 MB, 5.56771 sec, 120.696 MB/sec
46: ArrayDevice::ReadIntoCache: 22 pages, 704 MB, 5.62876 sec, 125.072 MB/sec
@\vdots@
\end{lstlisting}
% % \end{verbatim}
% % \endgroup
% \caption{
% Time measurements for a few of the 96
% {\tt ArrayDevice}
% processes
% in a typical time step.
% }
% \label{fig:ArrayDeviceReadIntoCacheTime}
% \end{figure}
% Time measurements for a few of the 96
% {\tt ArrayDevice}
% processes
% in a typical time step.
% The throughput is about what you'd expect from these hard drives.
The reading throughput is close to the maximal throughput for this type
of hard drives.
For every server the page reading commands are scheduled before the
page writing commands of the last phase of the iteration.
The writing of the pages will therefore start only after the completion
of the page read commands.
The clients write pages in parallel,
with the typical timing as follows:
% \begin{figure}
\begin{lstlisting}[escapechar=@,basicstyle=\scriptsize\ttfamily,frame=l]
8: Array::write 6144-6271 x 2432-2559 x 0-16383:: 4 GB, 22.3279 sec, 183.448 MB/sec
4: Array::write 3072-3199 x 2432-2559 x 0-16383:: 4 GB, 22.784 sec, 179.776 MB/sec
11: Array::write 8448-8575 x 2432-2559 x 0-16383:: 4 GB, 22.7165 sec, 180.31 MB/sec
12: Array::write 9216-9343 x 2432-2559 x 0-16383:: 4 GB, 23.083 sec, 177.447 MB/sec
15: Array::write 11520-11647 x 2432-2559 x 0-16383:: 4 GB, 23.0278 sec, 177.872 MB/sec
7: Array::write 5376-5503 x 2432-2559 x 0-16383:: 4 GB, 23.2437 sec, 176.22 MB/sec
0: Array::write 0-127 x 2432-2559 x 0-16383:: 4 GB, 23.7938 sec, 172.145 MB/sec
14: Array::write 10752-10879 x 2432-2559 x 0-16383:: 4 GB, 24.5307 sec, 166.975 MB/sec
10: Array::write 7680-7807 x 2432-2559 x 0-16383:: 4 GB, 24.2152 sec, 169.15 MB/sec
6: Array::write 4608-4735 x 2432-2559 x 0-16383:: 4 GB, 24.6648 sec, 166.066 MB/sec
2: Array::write 1536-1663 x 2432-2559 x 0-16383:: 4 GB, 24.7291 sec, 165.635 MB/sec
3: Array::write 2304-2431 x 2432-2559 x 0-16383:: 4 GB, 25.3921 sec, 161.31 MB/sec
9: Array::write 6912-7039 x 2432-2559 x 0-16383:: 4 GB, 21.8906 sec, 187.113 MB/sec
13: Array::write 9984-10111 x 2432-2559 x 0-16383:: 4 GB, 22.0268 sec, 185.956 MB/sec
1: Array::write 768-895 x 2432-2559 x 0-16383:: 4 GB, 21.7936 sec, 187.945 MB/sec
5: Array::write 3840-3967 x 2432-2559 x 0-16383:: 4 GB, 22.2789 sec, 183.851 MB/sec
\end{lstlisting}
% \caption{
% Time measurements of the parallel execution of
% {\tt Array::write}
% by the 16 client processes in a typical time step.
% }
% \label{fig:writeTime}
% \end{figure}
The results for page writing are fairly uniform,
with the total time for each process between 22 and 25.5 seconds.
The aggregate througfhput for this phase is therefore over 2.5 GB/sec.
We did not implement explicit caching for writing pages.
% This is obtained without a caching mechanism in the server.
% The {\tt Array:write} method transmits pages to the server over the network
% and
% For page writing
% we did not implement a caching mechanism in the server
% because 
% % the aggregate throughput is very high and 
% the expected
% additional improvement from caching would not be substantial.
The implementation of {\tt Array::write}
is analogous to the implementation of {\tt Array::ReadIntoServerCache}
(see section \ref{sec:caching}):
\begin{lstlisting}[escapechar=@]
void Array::write(Domain * domain)
{
	for @{\em every}@ page @{\em in}@ domain
		start @{\em write}@ page @{\em to the appropriate server}@
}
\end{lstlisting}
There is
a limited caching effect as a result of the {\tt \color{blue} start} command:
having transmitted the page to the server, the client disconnects
and proceeds to transmit pages to other servers, while the server starts
writing the page to disk only after the client has disconnected.
We found that writing pages to hard drive with the {\tt O\_DIRECT} flag
is about twice as fast as reading, presumably because of buffering.
The typical throughput measured was between 230 and 240 MB/sec:
% 53: ArrayDevice::write: 1 pages, 32 MB, 0.136526 sec, 234.387 MB/sec
% 77: ArrayDevice::write: 1 pages, 32 MB, 0.136314 sec, 234.752 MB/sec
% 86: ArrayDevice::write: 1 pages, 32 MB, 0.134568 sec, 237.798 MB/sec
% 30: ArrayDevice::write: 1 pages, 32 MB, 0.136555 sec, 234.338 MB/sec
% 7: ArrayDevice::write: 1 pages, 32 MB, 0.139071 sec, 230.098 MB/sec
% 94: ArrayDevice::write: 1 pages, 32 MB, 0.132235 sec, 241.993 MB/sec
% 46: ArrayDevice::write: 1 pages, 32 MB, 0.136529 sec, 234.382 MB/sec
% 25: ArrayDevice::write: 1 pages, 32 MB, 0.137174 sec, 233.281 MB/sec
% 26: ArrayDevice::write: 1 pages, 32 MB, 0.138168 sec, 231.602 MB/sec
% 23: ArrayDevice::write: 1 pages, 32 MB, 0.138837 sec, 230.486 MB/sec
% 95: ArrayDevice::write: 1 pages, 32 MB, 0.134038 sec, 238.738 MB/sec
% 74: ArrayDevice::write: 1 pages, 32 MB, 0.133987 sec, 238.829 MB/sec
% 65: ArrayDevice::write: 1 pages, 32 MB, 0.133562 sec, 239.589 MB/sec
% \begin{figure}
\begin{lstlisting}[escapechar=@,basicstyle=\scriptsize\ttfamily,frame=l]
@\vdots@
22: ArrayDevice::write: 1 page, 32 MB, 0.136958 sec, 233.648 MB/sec
78: ArrayDevice::write: 1 page, 32 MB, 0.135615 sec, 235.962 MB/sec
73: ArrayDevice::write: 1 page, 32 MB, 0.13452 sec, 237.883 MB/sec
64: ArrayDevice::write: 1 page, 32 MB, 0.136484 sec, 234.46 MB/sec
8: ArrayDevice::write: 1 page, 32 MB, 0.137042 sec, 233.505 MB/sec
54: ArrayDevice::write: 1 page, 32 MB, 0.136533 sec, 234.376 MB/sec
87: ArrayDevice::write: 1 page, 32 MB, 0.136293 sec, 234.788 MB/sec
@\vdots@
\end{lstlisting}
% \caption{
% Time measurements of 
% {\tt ArrayDevice::write}
% for several single page write...........
% % for a few of the 96 servers processes
% in a typical time step.
% }
% \label{fig:writeTime}
% \end{figure}

% The wall clock time for the iteration 
% was measured using {\tt MPI\_Wtime}.

% measurements of page transmission times.
% writing a domain.
% reading a domain;
% reading a domain into cache;
% throughputs;
% number of parallel clients.
% some discussion of scalability; resource utilization.

% we did not optimie for performance -- it was far more important
% to show that we can generate the high level cde quickly, and that
% this code performs quite well;
%%% 0: <19> Total time(FFT 16 page lines) = 69.1438 sec

The conclusion of our performance analysis is that the presented
computation could be sped up by 25\% or more,
but greater benefits can be derived from a more balanced hardware configuration.
The aggregate throughput of the 24 hard drives of a cluster node is about 3 GB/sec,
about 3 times the capacity of the incoming network connection.
Furthermore, the FFTW computation takes only 10-20\% of the total iteration
time.
It appears that a 2--4-fold  increase in the network capacity
of the present cluster is likely
to result in a more balanced system with better hardware utilization,
and a total runtime of under 20 seconds per iteration.

\section{
Discussion and 
Conclusions
}
\label{sec:Conclusion}

In this paper 
we introduced process-oriented programming
as a natural extension of object-oriented programming for parallel computing.
We implemented a prototype of the process framework
and carried out a data-intensive computation.
We have shown that a complex and efficient application can be built
using only a few hundred lines of process-oriented code,
% which translate to
which is
equivalent to
many thousands of lines of object-oriented code with MPI.
The process-oriented code in this paper is an extension of C++,
but processes can be introduced into any object-oriented language.
% by a suitable compiler extension.
%
% Programmers are typically reluctant to learn a new programming language,
% and process-oriented programming uses almost no new syntax.
The syntax extension is minimal.
In C++, for example, it
amounts to adding a parameter
to the {\tt \color{blue} new} operator and
introducing the keyword
{\tt \color{blue} start}.
% Creating a process can be thought of as simply placing an object
% on a remote machine, and
% existing code can be easily modified to run in parallel.
Combined with the fact that
% The minimal syntax extension and the fact that 
creating a process can be thought of as simply placing an object
on a remote machine,
it
suggests that
a lot of
existing code can be easily modified to run in parallel.
Potentially the most important impact of the process-oriented extension of 
languages, such as C++ and Python, is a widespread adoption of parallel
programming,
% as the simple syntax of process creation should also encourage
% the use of objects/processes instead of threads.
as application developers
realize the ability
to easily create processes instead of using thread libraries
and
to place different
objects on different CPU cores.
% The simple syntax of process creation should also encourage 
% the use of objects/processes instead of threads.

The process-oriented programming model 
is based on a simple hardware abstraction:
the computer consists
of a collection of processors, interconnected by a network,
% with each processor being able to run a single process.
where each processor is capable of running multiple processes.
The run-time system is responsible for mapping the abstract model
onto a concrete hardware system,
and must provide the programmer with system functions describing the
state of the hardware.
The hardware abstraction of the process-oriented model makes
it possible to create portable parallel applications
and applications that run in the cloud.
% The run-time system should contain functions
% describing the available hardware resources.
% While process-oriented programming can be used to create
% applications for a specific hardware architecture,
% it can also be used to create portable parallel applications
% and applications that run in the cloud.
% By avoiding hard-coded machine names,
% process-oriented programs can be made portable.

Processes are accessible by remote pointers.
Syntactically, executing a method on a remote process is not different
from method execution on an object.
Any class of an object-oriented language can be interpreted as a process,
but even more importantly, in the process-oriented framework
only processes that are class instances
% implement classes
are allowed.
% Objects can be interpreted as processes, but even more important is
% the dual point of view: the programmer is restricted to processes
% that represent objects.
% This is a natural kind of parallelism, i.e. something that people find
% easy to reason about.
% unlike
% unrestricted shared memory -- violates encapsulation;
% here processes help by providing controlled access to shared memory.
% We argue that {\em object-based parallelism}
% is a more natural way to reason about 
% parallelism than shared memory or message passing,
% yet process-oriented programming is more general, and includes these
% frameworks as special cases.
We argue that {\em object-based parallelism}
is a high level abstraction, which is
naturally suitable for reasoning about parallelism.
Although shared memory and message passing can be realized
in a process-oriented language,
these are lower implementation-level models.
% We argue that this is a more natural framework for reasoning about
% parallelism than shared memory or message passing,
% yet process-oriented programming is more general, and includes these
% frameworks as special cases.
%
Process inheritance
is a powerful aspect of object-based parallelism,
as it enables the definition of new processes
% makes possible to define new processes
in terms of previously defined processes.
In combination with process pointers, it gives the programmer
the flexibility
to adapt the computation to the hardware
% by assigning computational tasks to the available processors
by adding simple methods to class definitions
(see the comments at the end of section \ref{sec:TheArray} about
the computation of array page transpose).

A process-oriented program, like a typical sequential program
starts with a single main process.
The main process may launch new processes on remote machines,
as easily as it can create objects.
In contrast with MPI,
% process management is in the responsibility of the programmer.
processes are explicitly managed by the programmer.
Process launching is part of the program,
and is not determined by the runtime command line parameters.
% (Although MPI provides process process spawning functions, the arguments
% provided in the {\tt MPI\_Info} object are not standardized.)
Using process pointers and language data structures,
the programmer can form groups of processes, assign process ids
and perform tasks that in MPI would require using communicators.

Processes exchange information by executing remote methods,
rather than via shared memory or message passing.
We'd like to use the analogy that writing programs with message passing
today is like writing programs with GOTOs fifty years ago:
it is easy to write intractable code.
And just like GOTO statements are used in assembly languages,
message passing is a low-level language construct
underlying remote method execution in process-oriented programming.
% in parallel programming.
% The process framework is not shared memory, it is not PGAS, and it is
% not distributed programming. Surely, people have built programs
% as collections of processes before, but they used message passing
% to exchange information between processes, and they rarely thought
% of their processes as representing objects.

% By default, remote method execution, like any statement in a sequential
% program, must complete before the following statement is executed.
% As an extension of a sequential language, 
% each statement of a process-oriented language must complete
% before the execution of the next statement starts.
We introduced the {\tt \color{blue} start} keyword
to enable parallel execution of remote methods.
In order to use the keyword the programmer must decide whether
there is a need to wait for the remote task to complete before
proceeding with the computation.
In general, this decision is easy and intuitive,
but keeping track of task dependencies is not.
% In a typical large computation many tasks will be started and it could
% be difficult to keep of task dependencies.
Each process executes only one method at a time
and remote method execution requests are queued.
The programmer must keep in mind the state of the execution queue
for every process.
This is possible only for very simple scenarios.
We used barriers to synchronize processes.
Barrier synchronization helps the programmer to keep track of
the execution queues of the processes,
but it may reduce the parallelism of the computation.

% Inheritance for processes:
% it is possible to define processes in terms of previously defined processes.
% Also, the derived class process can be launched alongside the base class
% process with essentially no overhead (when launched on the same machine).
% Remote pointers make it possible to either bring computation to data
% or the other way around.
% I.e. in combination with inheritance it is easy to put computation
% where you want it, or... computation/communication allocation
% Bring the data to program, bring the program to data.
% The decision here is similar to dereferencing a pointer.
% Programmers are aware of the complexity of data access
% via additional pointer indirection.
% Process framework is adaptable to architecture:
% we can schedule more processes where we have more CPU cores, etc.,
% and we can easily shift the computation there.
% example: transpose can be done either on the server or on the client.

We view a large data object as a collection of persistent processes.
For a large data object a negligible amount of additional storage
space is needed to store serialized processes alongside the data.
Process persistence is needed to enable stopping and restarting
a computation,
and to make a data set accessible to several simultaneous applications.
It is also needed to develop basic mechanisms for fault tolerance.
In this paper we showed that the process-oriented view of a large data object 
is very powerful:
using only a small amount of code the programmer can copy and reformat 
very large data objects, and even carry out complex operations,
such as the Fourier transform.
Yet, software users and application developers tend to see
% Unfortunately, application developers and software users tend to see
a data set as consisting of ``just data'', 
and being independent of a specific programming language.
We stop short of suggesting a solution for this problem,
but in this context it is worth recalling 
the CORBA standard 
\cite{Henning:2006:RFC:1142031.1142044}.
% was developed to bridge the language problem.

The introduction of process-oriented programming was motivated by
our research in data-intensive computing.
The data-intensive Fourier transform
computation was carried out on a small cluster
of 8 nodes (96 cores).
Our measurements indicate that Petascale
data-intensive computations 
can be efficiently carried out on a larger cluster,
with more nodes and significantly increased network bandwidth.
% throughput capacity.
%
Such a cluster can serve as a general-purpose data-intensive computer,
whose operating system and applications are developed
% within the process-oriented framework.
as process-oriented programs.
% The process framework can be used to develop
% an operating system for a data-intensive computer.

% An implementation of object oriented processes would provide a programmer with
% the ability to quickly produce code for complex computations
% with large data sets.
% We demonstrated the efficiency of this framework in a large-scale computation
% of a three-dimensional Fourier transform.
% Using the techniques presented here (and similar) such code
% can be made very efficient
% (achieve very high rates of data parallelism) for a wide range of problems.

% syntax should make it easy to replace
% threads, operating systems;
% programming multi-core processors;
% comments on threads, GPUs, Tile processors, etc.
% Tools....
% MapReduce becomes trivial in the current framework.
% MapReduce and DBMS is here
% \cite{pavlo2009comparison}
% cloud

The introduction of process-oriented programming in this paper is far from
complete.
It is merely the first step of an extensive research program.
% Our prototype implementation of the process framework enabled us
% to test a subset of the features described in this paper.
We tested a substantial subset of the 
process framework
% features of the process framework
in a prototype implementation.
A full-fledged implementation
% of the process framework 
must include a compiler and a run-time system
that substantially expand the basic prototype.

\section{
Acknowledgements
}
% \begin{acks}
% Acknowledgements
I am grateful to J. J. Bunn for discussions that
significantly improved the presentation of the material.
% \end{acks}

% This research was partially funded by a grant from Intel Corp.

% maybe this goes in a footnote....
% \newpage

\bibliographystyle{abbrv}
\bibliography{paper}

\end{document}